# THE UNIDENTIFIED GALACTIC EGRET SOURCES

Philip Kaaret

*Columbia University, New York, NY 10027, USA*

ABSTRACT

The nature of the unidentified gamma-ray point sources in the galactic plane is a long standing puzzle of gamma-ray astronomy. Kaaret and Cottam (1996) showed that 16 of the 25 unidentified gamma-ray point sources near the galactic plane lie in or near associations of massive stars, likely sites of pulsar formation, and that the intrinsic luminosity distribution of the sources is consistent with that of known gamma-ray pulsars. Here, I show there are approximately 5-7 times as many visible gamma-ray pulsars as supernovae remnants. This strengthens the hypothesis that pulsars constitute the majority of the unidentified gamma-ray point sources near the galactic plane.

INTRODUCTION

The COS-B satellite, launched in 1975, discovered a population of unidentified gamma-ray point sources concentrated along the galactic plane (Swanenburg *et al.* 1981). Recently, EGRET surveyed the 0.03-10 GeV sky and found 129 point sources (Thompson *et al.* 1995). The majority of identified sources are AGN (von Montigny *et al.* 1993). The only galactic point sources with convincing identifications are pulsars. All six pulsars are young, less than $10^6$ years old, and nearby, closer than 3 kpc. There are 25 unidentified sources at low galactic latitudes, $|b| < 5°$. The problem of the unidentified galactic gamma-ray point sources still remains.

That the only identified galactic gamma-ray point sources are pulsars suggests that the unidentified galactic sources may also be pulsars (Halpern and Ruderman, 1993; Helfand, 1994). The progenitors of pulsars are massive stars which are often found clustered in OB associations. The lifetimes of OB associations are typically a few times the $10^7$ year lifetimes of massive stars, and are significantly longer than the $10^6$ year lifetimes of gamma-ray pulsars. It should, therefore, be possible to use OB associations to trace the population of young pulsars.

Kaaret and Cottam (1996) compared the positions of the sources in the second EGRET catalog to the catalog of OB associations produced by Mel'nik and Efremov (1995), see Figure 1. Among the unidentified EGRET sources, there are 9 sources with error boxes within or overlapping OB association boundaries, and 7 sources within 1° of an OB association. All 16 sources are at low galactic latitudes, $|b| < 5°$. The probability that the superposition of EGRET sources with OB associations occurred by chance is low, $1.0 \times 10^{-4}$. Correcting for chance superpositions, 13±2 of the superpositions are genuine. The intrinsic luminosity distribution of the 16 sources, obtained using the known OB associations distances, is consistent with the luminosity distribution of known gamma-ray pulsars. The correlation of unidentified EGRET sources with OB associations and the similarity of their luminosity distributions to that of pulsars suggests that a significant fraction of the unidentified sources are gamma-ray pulsars.

NATURE OF THE GALACTIC GAMMA-RAY SOURCES

The correlation of EGRET sources with OB associations indicates that the galactic gamma-ray sources are members of a young stellar population. The absence of gamma-ray sources in globular clusters (Michelson *et al.*, 1994) reinforces this interpretation. There are several classes of potential gamma-ray emitting objects associated with young stellar populations including young pulsars, supernovae remnants (SNR), and supernovae remnants interacting with molecular clouds in OB associations (SNOB). Pulsars are the only sources with high confidence detections in gamma-rays. However, it is commonly thought that interactions of cosmic-rays accelerated in supernovae remnants should produce observable gamma-ray fluxes (Pinkau, 1970).

Montemerle (1979) found a correlation between COS-B sources and OB associations which he attributed to SNR interacting with dense matter in the associations. SNR, from the catalog of Green (1988), overlap the error boxes of 7 of the 16 sources near OB associations. Three of these, IC443, $\gamma$-Cygni, and W28, are good candidates for cosmic-ray induced gamma-ray emission as they show high-velocity shocks and interaction with adjacent molecular clouds. These three sources were originally identified by Montemerle (1979) as likely gamma-ray luminous SNR. Sturner and Dermer (1995) found a significant correlation between SNR and sources in the first EGRET catalog. However, the correlation did not persist with the second EGRET catalog (Sturner, Dermer, and Mattox, 1996).

Assuming similar birth rates and luminosities for pulsars and SNR, we can estimate the ratio of gamma-ray visible pulsars to SNR as $f_{PSR}\, \tau_{PSR}/f_{SNR}\, \tau_{SNR}$, where $f$ is the beaming fraction and $\tau$ is the lifetime over which the source emits gamma-rays for pulsars (PSR) and supernovae remnants (SNR). The average age of the seven known gamma-ray pulsars is $1.6 \times 10^5$ years. The pulsar beaming fraction is uncertain, but is likely to be high as 5 of the 6 pulsars with the highest values of $\dot{E}/d^2$ are seen in gamma-rays. We assume a beaming fraction of 25%. The average age of the SNR identified by Esposito et al. (1996) as gamma-ray luminous is 7000 years. The beaming fraction for SNR is unity. The ratio of gamma-ray visible pulsars to SNR is then 6. If all of the unidentified Galactic EGRET sources are either pulsars or SNR, there should be 3-5 gamma-ray SNR. This is consistent with the number of sources identified by Esposito et al. (1996) and with the number of interacting SNR found above. Because pulsars emit gamma-rays over a significantly longer lifetime than SNR, the number of gamma-ray pulsars is significantly larger than the number of gamma-ray SNR.

We note that our technique of identifying gamma-ray sources does not distinguish between pulsed emission and nebular emission powered by a pulsar wind. As there is confirmed high energy gamma-ray nebular emission only from the Crab, it seems that pulsed emission is more common than nebular emission. This is consistent with theoretical models of the nebular emission (Harding, 1996). However, pulsars within OB associations may be located in regions of significantly enhanced soft photon density. These soft photons could be inverse-Compton scattered by the pulsar wind and may produce detectable emission near 100 MeV.

## NUMBER OF GAMMA-RAY PULSARS

To estimate the total population of visible gamma-ray pulsars we must correct for the fraction of pulsars expected to lie outside OB associations either because they are born outside an OB association or because their velocity carries them away from their parent association after birth. In addition, account should be taken that some of the gamma-ray sources in OB associations are SNR rather than pulsars. Following the discussion above, we assume that four of the sources coincident with OB associations are SNR. Using the estimate of the fraction of field OB stars of 0.5 given by Humphreys and McElroy (1984), the velocity distribution from Lyne and Lorimer (1994), and assuming a typical gamma-ray pulsar age of $2\times10^5$ years, gives an estimated total number of visible gamma-ray pulsars of $27\pm7$; this total includes the known gamma-ray pulsars. There are approximately 20 pulsars among the 25 unidentified EGRET sources near the Galactic Plane, $|b| < 5°$. A similar analysis by Yadigaroglu and Romani (1996) supports this assertion.

This estimate of the number of visible gamma-ray pulsars can be compared to the pulsar birth rate. The local pulsar birth rate is $2\times10^{-5}$ pulsars kpc$^{-2}$ yr$^{-1}$ (Lyne and Graham-Smith, 1990). Adopting an average pulsar age of $2\times10^5$ years and a distance limit of 3 kpc corresponding to the sensitivity of EGRET gives 110 pulsars. Of these, $27/110 = 0.24$ are visible in gamma-rays. Within errors, this estimate of the gamma-ray beaming fraction is consistent with the predictions of Yadigaroglu and Romani (1995) based on a pulsar magnetospheric outer-gap model.

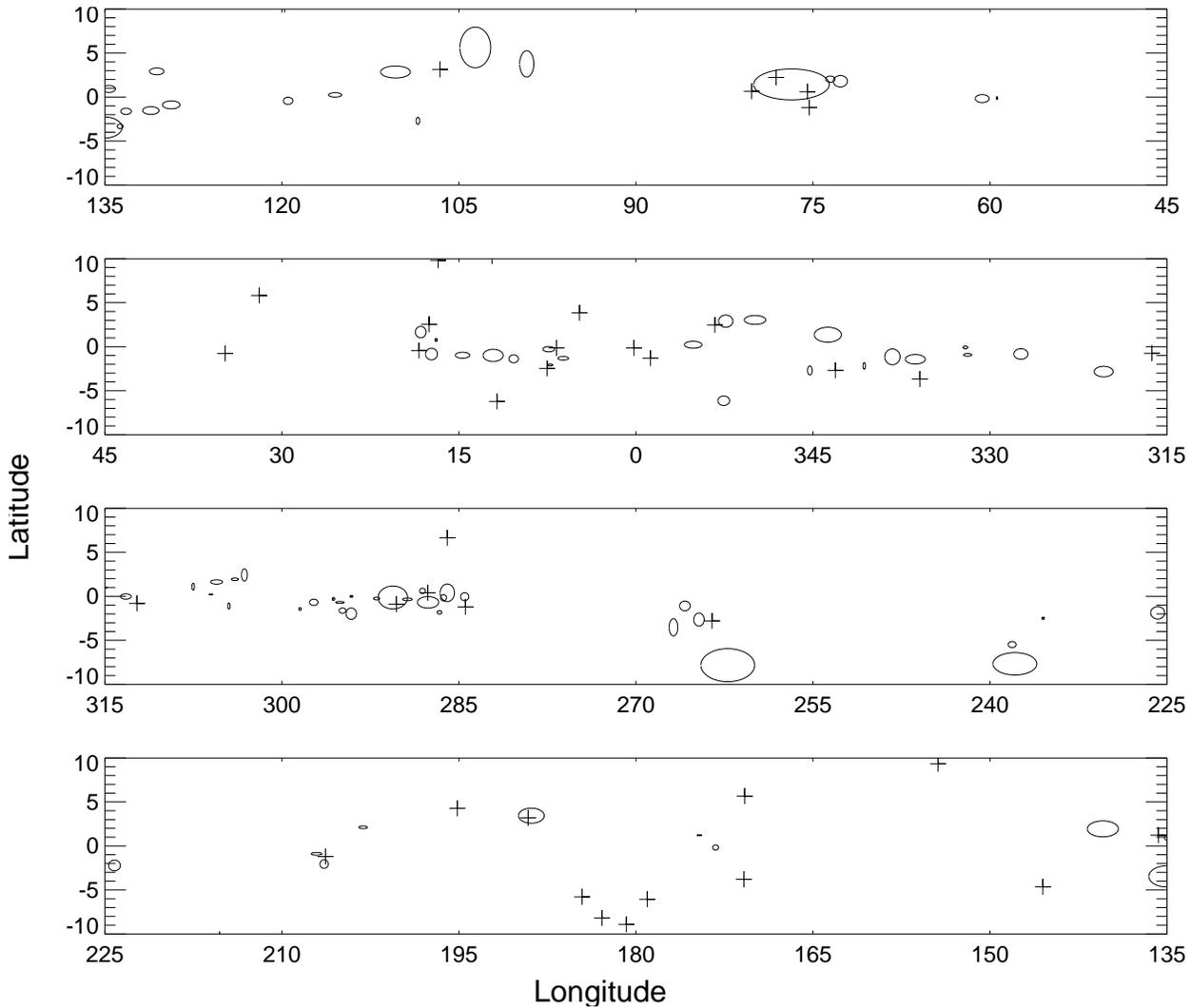

Fig. 1. Positions of EGRET sources and OB associations along the Galactic plane. The crosses are EGRET sources (identified and unidentified). The ellipses show the location and size of the OB assocations.

The fraction of gamma-ray pulsars visible in the radio, $6/(27\pm7) = 0.22\pm0.08$, is consistent with the radio beaming fraction of 0.2 if there is no alignment between the radio and gamma-ray beams. The total number of gamma-ray pulsars and the fraction visible in the radio is consistent with the predictions of an outer-gap model (Yadigaroglu and Romani, 1995), but is inconsistent with the predictions based on a polar-cap model (Sturner and Dermer, 1996).

PROSPECTS FOR FURTHER IDENTIFICATION

The most convincing identification of individual sources as pulsars would come from detection of pulsations at gamma-ray or other wavelengths. Direct pulsation searches of the gamma-ray data are hampered by the extremely low photon flux and are unlikely to yield results except, perhaps, for the brightest sources. While many of the gamma-ray pulsars should be radio quiet, pulsation searches in the radio may detect a few sources. It is important that these searches be extended to high dispersions as the environment of OB associations may produce high dispersion.

It may be possible to identify some of the sources from their soft x-ray emission. For sources located within OB associations, the soft x-ray flux may be obscured by material within the association. Notwithstanding, a ROSAT counterpart was found for the EGRET source 2EG2020+4026 located in the Cygnus 1,8,9 association (Brazier, 1996). The properties of this counterpart are suggestive of soft x-ray emission from a neutron star surface. However, the flux is too weak to permit a pulsation search in x-rays. Unfortunately, the ROSAT PSPC is no longer in operation and the sensitivity of the ROSAT HRI is significantly worse than that of the PSPC. Searches are possible with ASCA, however, the effective area of ASCA is limited at low energies where the majority of the x-ray flux is produced. The best avenues to pursue for detection of x-ray counterparts are a search of the ROSAT pointed archive and observations with future instruments with high effective area at low energies such as Spectrum-X-Gamma (Schnopper, 1990).

The gamma-ray spectra of the sources may also give some information about their identity. Approximately one third of the unidentified galactic EGRET sources have hard spectra suggestive of those observed from known gamma-ray pulsars. For many sources, the gamma-ray flux is too low to obtain accurate spectra. In addition, spectral analysis of sources in crowded fields with EGRET is complicated due to the energy dependence of the point spread function. Structure in the diffuse emission tends to produce higher flux estimates at lower energies where the point spread function is larger. This may produce softer spectra for sources in complicated regions such as OB associations.

CONCLUSION

The majority of the 25 unidentified gamma-ray sources near the galactic plane are most likely pulsars with approximately 3-5 gamma-ray loud SNR. In addition, the isotropic population of unidentified sources, most likely AGN, and the population of intermediate-latitude sources (Grenier, 1995; Mukherjee et al., 1995) should contribute 1-4 low-latitude sources . A population of several sources of unknown origin cannot be excluded. We note that when the sources coincident with OB associations are removed, a concentration of 4 sources towards the inner part of the Galaxy remains. These sources may be very young (hundreds of years old) and luminous pulsars or a new class of gamma-ray emitting object (Pohl 1996). If these sources are young pulsars similar to the Crab, they should emit radiation over a very broad band. Therefore, it may prove fruitful to conduct deep searches for pulsations within the error boxes of these four sources both in the radio, at high dispersion measures, and in x-rays.